\documentclass[seceq]{ptptex}

\usepackage{graphicx}
\usepackage{wrapft}

\newcommand{\be}{\begin{equation}} \newcommand{\ee}{\end{equation}}
\newcommand{\bea}{\begin{eqnarray}} \newcommand{\eea}{\end{eqnarray}}
\newcommand{\s}{\sigma}

\def\(({\left(} \def\)){\right)}

\title{% %You can use \\ for explicit line-break.
  Quantum Annealing of Hard Problems}

\author{% %Use \scshape for the family name.
  Thomas \textsc{J\"org$^1$}, Florent \textsc{Krzakala$^{2,3}$},
  \underline{Jorge \textsc{Kurchan$^4$}} and A. C
  . \textsc{Maggs$^{2}$}

}

\inst{% %Affiliation, neglected when [addenda] or [errata].
  $^1$ CNRS et ENS UMR 8549, 24 Rue Lhomond, 75005
  Paris, France, LPTENS\\
  $^2$ CNRS; ESPCI, 10 rue Vauquelin, Gulliver, Paris, France
  75005, PCT\\
  $^3$ Theoretical Division and Center for Nonlinear Studies, Los
  Alamos National Laboratory, NM 87545 USA \\
  $^4$ CNRS; ESPCI, 10 rue Vauquelin, UMR 7636, Paris, France 75005,
  PMMH }

\abst{% %This abstract is neglected when [addenda] or [errata].
  Quantum annealing is analogous to simulated annealing with a
  tunneling mechanism substituting for thermal activation.  Its
  performance has been tested in numerical simulation with mixed
  conclusions. There is a class of optimization problems for which the
  efficiency can be studied analytically using techniques based on the
  statistical mechanics of spin glasses. }

\begin{document}

\maketitle

\section{Introduction}

Solving hard combinatorial problems by temperature annealing is a
classic method in computer science \cite{Scott}. The problem is
formulated in terms of a cost function, which one can identify as an
energy.  The dynamics are then a combination of a systematic descent
in energy perturbed by a 'noise', which is turned off gradually,
inducing occasional upward jumps. This {\em thermal activation} allows
the system to surmount barriers and avoid getting blocked in a local
minimum.

Quantum annealing is an analogous procedure, but with {\em quantum
  tunneling} substituting for thermal activation.  A major question is
whether quantum algorithms that anneal by gradually turning off the
term that induces quantum jumps \cite{QA,GF} -- such as a transverse
magnetic field in spin systems -- can be an efficient strategy that
outperforms classical optimization methods.

In the past few years, there have been several attempts to simulate
quantum annealing of simple "benchmark" models, in order to estimate
the scaling of the time $\tau$ needed in terms of the system size
$N$~\cite{QA,GF,Young}.  This time is bounded by the minimal energy
gap $\Delta$ between the ground state and the first excited state: one
can estimate~\cite{QA,GF,GAP} the time needed for quantum annealing to
find the ground state as $\tau \propto \Delta^{-2}$ .  In the most
optimistic scenario \cite{Young}, a problem taking $\tau \sim e^{a N}$
in a classical computer would be solved in $\tau \sim N^\alpha$ in a
quantum one -- corresponding to a minimal gap $~\sim N^{-\alpha/2}$.
A more modest achievement, would be that the problem stays
exponential, but with a smaller coefficient $a$ (as is the case of the
Grover~\cite{Grover} algorithm): this is what would at best happen if
the gap is exponentially small.

Because, at present, the simulations are done emulating quantum
systems with classical computers, the sizes accessible are rather
small, and the extrapolation of the gap dependence to larger systems
risky.  For this reason, it is particularly desirable to have an
analytic framework within which one can compute $\Delta$ for a
representative class of models.  In a recent work, we have considered
the group of systems having a 'Random First Order' (RFO) glass
transition~\cite{JKKM}.  This class of models \cite{PSPIN} is
currently believed to be the mean-field version of the glass
transition \cite{GLASS} and random heteropolymer folding
\cite{POLYMERS}. It also includes random constraint satisfaction
problems such as random satisfiability \cite{SAT} (the energy to be
minimized is the number of violated constraints) and is closely
related to the random code ensemble in coding theory \cite{BOOK}.

The phase diagrams of quantum spin glasses have been investigated
extensively over the last thirty years, using a formalism combining
the Replica~\cite{MPV} and the Suzuki-Trotter
methods~\cite{Goldschmidt,Obuchi,QREM-finitep} in order to deal
simultaneously with disorder and quantum mechanics.  This formalism
was first applied by Goldschmidt~\cite{Goldschmidt} to the paradigm of
the class, the quantum Random Energy Model.  An important result is
that the quantum transition is of first order (with a discontinuity in
the energy) at low temperature for all RFO models, unlike their
classical transition in temperature, which is thermodynamically of
second order.

In order to assess the efficiency of quantum annealing knowledge of
the thermodynamics of the model is not enough; we must compute the gap
between the two lowest eigenvalues. This is not given by the usual
replica solution, and one has to develop stronger techniques to obtain
it.  In what follows, we shall first show how to solve the Quantum
Random Energy Model by elementary methods, including the calculation
of the gap. Quantum Annealing turns out not to be an efficient way of
computing its ground state for reasons that are very clear and require
minimal technique to understand.  Unfortunately, such elementary
perturbative methods do not allow us to solve a general model of the
class, and in particular to compute the minimal gap. We have hence to
resort to more powerful, yet less transparent, methods which we have
developed and we shall describe below.

As we shall see, the minimal gap between the lowest states in all
models of the RFO class can be expected to be exponentially small in
$N$, implying that quantum annealing is an exponentially slow
algorithm for those problems.

\section{The simplest case: Random Energy Model}

The random energy model \cite{REM} is the simplest mean field spin-glass
model \cite{PSPIN}, yet it allows us to understand the very complex
mechanisms of the mean-field spin-glass \cite{PSPIN} and glass
transitions \cite{GLASS}. We shall first use this model to study the
effect of a quantum transverse field in the mean field transition in
spin-glasses. The problem was already investigated twenty years
ago \cite{Goldschmidt}, by considering the $p \to \infty$ limit of
p-spin model using the replica method and the Suzuki-Trotter
formalism, and many generalizations followed\cite{Obuchi}. Here we
shall show that it can be solved in a very simple way, that does not
require the use of replica techniques.  We consider $N$ quantum spins
$\sigma$ in a transverse field $\Gamma$ with the Hamiltonian
\begin{equation} 
{\cal{H}}(\{{\sigma \}}) =  {\cal}E(\{{\sigma^z\}}) + \Gamma V =
{\cal}E(\{{\sigma^z\}}) + \Gamma \sum_{i=1}^N \sigma^x_{i} 
\label{Hamiltonian}
\end{equation}
${\cal}E(\{{\sigma^z\}})$ denotes a quenched random function that is
diagonal in the basis $|\sigma^z_1 \rangle \otimes |\sigma^z_2 \rangle
\otimes ... \otimes |\sigma^z_N\rangle $ and consists of $2^N$ random,
uncorrelated values.  These values are taken from a Gaussian
distribution of zero mean and variance $N/2$, as in the classical REM
\cite{REM}.  ${\cal{H}}$ is thus a $2^N \times 2^N$ matrix whose
entries are ${\cal{H}}_{aa}=E_{a}$ (the random energies) and
${\cal{H}}_{ ab} = \Gamma$ if $a$ and $b$ are two configurations that
differ by a single spin flip, and zero otherwise. Note that since
${\cal{H}}$ is sparse, it can be studied numerically even when its
dimensions  are large.

Just as in the classical case, a concrete implementation of the model
is a spin glass with $p$-spin interactions, in the
large $p$ limit:\\
\begin{equation}
H=-\sum_{i_{1}<\ldots <i_{p}} J_{i_{1}\ldots i_{p}}
\sigma_{i_{1}}^{z} \ldots \sigma_{i_{p}}^{z} -\Gamma
\sum_{i=1}^{N} \sigma_{i}^{x} \equiv T+\Gamma V,
\label{pspin}
\end{equation}
where $J_{i_{1}\ldots i_{p}}$ are Gaussian variables with  variance $\sqrt{\frac{p!}{N^{p-1}}}$ 

\subsection{Two easy limits}\hspace{-0.4cm} The model is trivially solved in the
limits  $\Gamma \to 0$ and $\Gamma \to \infty$. 

{\em a) $\Gamma=0$: \;\;\;\;\;}For $\Gamma=0$, we
recover the classical REM with $N$ Ising spins and $2^N$
configurations, each corresponding to an energy level $E_{\alpha}$
\cite{PSPIN}: Call $n(E)$ the number of energy levels belonging to the
interval $(E,E+dE)$; its average over all realizations is easily
computed: $\overline{n(E)} = 2^N P(E) \propto
e^{N\left((\ln\!2-E^2/N^2\right))} = e^{Ns(E/N)}$, where \\$s(e)=\ln\!2-e^2$
(with $e=E/N$). There is therefore a critical energy density
\\$e_0=-\sqrt{\ln\!2}$ such that, if $e<e_0$, then with high probability
there are no configurations while if $e>e_0$ the entropy density is
finite. A transition between these two regimes arises at $\frac
1{T_c}=\frac{ds(e)}{de}\big|_{e_0}=2\sqrt{\ln\!2}$ and the
thermodynamic behavior follows: {\rm i)} For $T<T_c$,
$f_{\rm REM}=-\sqrt{\ln\!2}$ and the system is frozen in its lowest energy
states. Only a finite number of levels (and only the ground state at
$T=0$) contribute to the partition sum. The energy gap between them is
finite. (\rm ii) For $T>T_c$, $f_{\rm REM}=-\frac 1{4T} - T \ln\!2$;
exponentially many configurations contribute to the partition sum.

{\em b):\;\;\;\;\;}In the opposite case of $\Gamma \rightarrow \infty$, the REM contribution
to the energy can be neglected. In the $\sigma^x$ basis, we find $N$
independent classical spins in a field $\Gamma$; the entropy density
is just given by the log of a binomial distribution between 
$-\Gamma N$ and $+\Gamma N$ and the free-energy density is $f_{\rm para}=
-T\ln\!2-T\ln\!\left((\cosh\!\Gamma/T\right))$.

\subsection{Perturbation theory}\hspace{-0.4cm}

Between these two extreme limits, it turns out that nothing much
happens: the system is either in the "classical" or in the "extreme
quantum" phase, and it jumps suddenly -- in a first order fashion --
from one to the other.  This is important for us, since it means that
most of the effort in quantum annealing is in fact wasted in staying
on the same state, until suddenly the wave function projects onto the
exact solution, and then never changes again.  At low value of
$\Gamma$, the free-energy density is that of the classical REM, while
for larger values it jumps to the quantum paramagnetic (QP) value
$f_{\rm QP}$; a first-order transition separates the two different
behaviors at the value $\Gamma$ such that $f_{\rm REM}=f_{\rm QP}$
(see center panel of Fig.~\ref{levels}).

Let us see how this comes about in the restricted case of zero temperature, using 
%\corrout{the} 
Rayleigh-Schr\"odinger perturbation theory~\cite{RS}.  Consider the
set of eigenvalues $E_k$ and eigenvectors $|k \rangle$ of the
unperturbed REM, when $\Gamma=0$. The series for a given perturbed
eigenvalue $E_i(\Gamma)$ reads
\[ E_i(\Gamma)=E_i + \left\langle i \left| \sum_{n=0}^{\infty} \Gamma V \! \left[ \frac
{Q}{E_i-{\cal H}_{0}}(E_i-E_i(\Gamma)+\Gamma V) \right]^{\!n} \; \right|i \right\rangle,
\nonumber
%\label{total}
\]
where the projector $Q=\sum_{\substack{k \neq i}} |k \rangle \langle
k|$ so that
\begin{align}
  &E_i(\Gamma)=E_i + \Gamma V_{ii} + \sum_{k \neq i} \frac{\Gamma^2
    V_{ik}V_{ki} }{E_i-E_k} + \cdots \,.
  \label{serie1}
\end{align}
Since $V_{ij} \neq 0$ if and only if $i$ and $j$ are two configurations
that differ by a single spin flip, odd order terms do not contribute
in Eq.~(\ref{serie1}), as one requires an even number of flips to come
back to the initial configuration in the sums. 
If the starting energy $E_i$ is close to the ground state,  it is negative and of order $N$. On the other hand,
the vast majority of levels have energies $E_k \sim O(\sqrt{N})$ . Hence, all terms of the form
$(E_k-E_i)\sim O(N)$: the spin flips from the lowest levels induced by the $\sigma^x$ operators do not
connect the lowest states amongst themselves. 
One obtains, starting from an {\it
  low eigenvalue} [$-E_i=O(N)$], that
\begin{equation} 
  \nonumber 
  \sum_{k \neq i} \frac{V_{ik}^2}{E_i-E_k} = 
  \frac{1}{E_i} \sum_{k=1}^N \left(1+\frac{E_k}{E_i}+\cdots \right) = 
  \frac{N}{E_i} + O\left(\frac{1}{N}\right)\!,  
\end{equation}
where we have used that the $E_k$ are random and typically of order
$\sqrt{N}$. Higher $n$th orders are computed in the same
spirit and are found to be $O(N^{n/2-1})$. Therefore, to all (finite)
orders, we have for the energy density $\epsilon_i=\frac{E_i}{N}$:
\begin{equation}  \epsilon_i(\Gamma)= \epsilon_i + \frac{\Gamma^2}{N \epsilon_i} +...=  \epsilon_i +O\left(\frac{1}{N}\right) \label{REM_perturb} \!. 
\end{equation}
This analytic result compares well with  {a} numerical evaluation 
of the eigenvalues (left panel of Fig.~\ref{levels}). Note that the energy
density of all extensive levels is independent of $\Gamma$ to leading
order in $N$, as are hence $s(e)$ and $f(T)$.

A similar  expansion can also be performed  around the extreme quantum limit, using $\Gamma V$ as a starting
point and ${\cal H}_0$ as a perturbation. Consider the  {ground
  state with} eigenvalue  {$E_0(\Gamma)$ and the unperturbed
  ground state having  all spins aligned along the $x$-direction, with $E_0^V(\Gamma)=-\Gamma N$}. In the base 
%\corrout{$\mid n_i  \rangle$} 
corresponding to the eigenvalues of $\Gamma V$, we find
\[ 
 {E_0(\Gamma)} =  {E_0^V(\Gamma)} + \langle  {0} \mid
{\cal{H}}_0 \mid  {0} \rangle + \sum_{k \neq  {0}}
\frac{|\langle k \mid {\cal{H}}_0\mid  {0} \rangle| ^2
}{ {E_0^V(\Gamma)}- {E_k^V(\Gamma)}} + \cdots \,.
\]
The  {first-order} term gives $\sum_{ {\alpha=1}}^{2^N} E^{\rm REM}_ {\alpha}
 {|v_\alpha|^2}$. Since the energies of the REM are random and uncorrelated
with $v_ {\alpha}$ this sums to $O(\sqrt{N}2^{-N/2})$. For the
 {second-order} term, one finds
 {
  \begin{align}
    \sum_{k \neq 0} \frac{|\langle k \mid {\cal{H}}_0\mid 0 \rangle|^2
    }{E_0^V(\Gamma)-E_k^V(\Gamma)} &= \frac{1}{E_0^V(\Gamma)} \sum_{k\neq 0}  \frac{|\langle k \mid {\cal{H}}_0\mid 0 \rangle|^2 }{1-E_k^V(\Gamma)/E_0^V(\Gamma)}
    \nonumber \\
    \approx \frac{1}{E_0^V(\Gamma)} \langle 0
    \mid {\cal{H}}_0^2 \mid 0 \rangle &= \frac{N}{2 E_0^V(\Gamma)} +
    o(1).
  \end{align} 
} 
Subsequent terms are treated similarly and give vanishing corrections
so that $ {\epsilon_0}(\Gamma) = -\Gamma - \frac {1}{2N\Gamma} +
O(\frac{1}{N^2})$. Again this {derivation} holds for other states with
extensive energies {$-E_i^V(\Gamma)=O(N)$}, the only tricky point being the
degeneracy of the eigenvalues \cite{foot2}, and for these excited
eigenstates, the perturbation starting from the large $\Gamma$ phase
yields $ \epsilon_i(\Gamma) = \epsilon_i^V(\Gamma)-\frac 1{2N\Gamma} +
O(\frac{1}{N^2})$.  Just as in the opposite classical limit, to leading order in $N$, energy, entropy
and free-energy densities are not modified by the perturbation.

\begin{wrapfigure}{r}{7.0cm}
  \includegraphics[width=7cm]{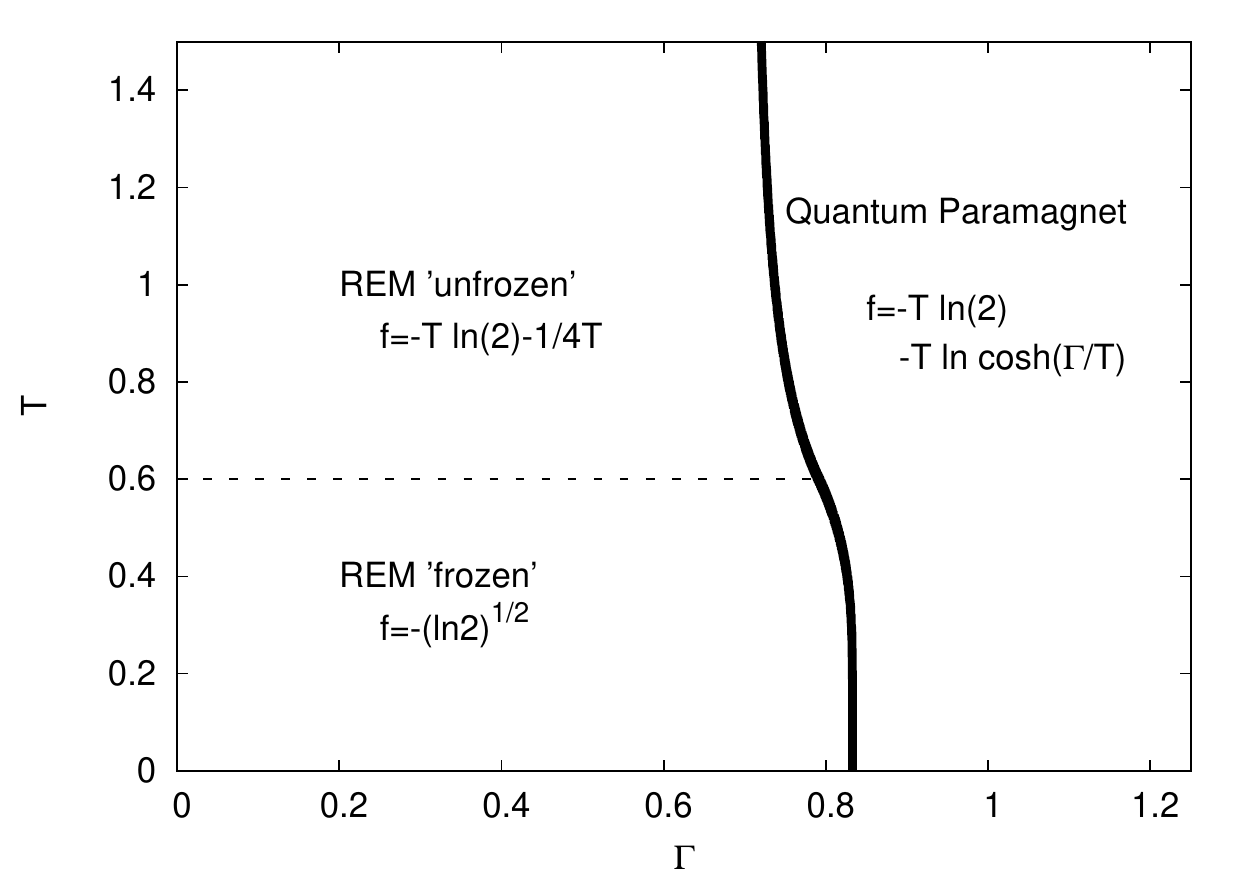}
\caption{Phase diagram of the QREM in temperature $T$ and transverse quantum
  field $\Gamma$. There are three phases: the classical unfrozen and frozen ones and the quantum one. The classical region is separated from the quantum one by a first order transition (thick line). At $T=0$ the quantum first order transition arises at  $\Gamma_c=\sqrt{\ln{2}}$ while the classical glass transition for
  $\Gamma=0$ is at $T_c=1/\sqrt{\ln{2}}$. 
  \label{phase}
}
\end{wrapfigure}

\subsection{Phase diagram and the closure of the gap}
All that we have been saying up to now implies that, to leading order,
the partition function can be written as:
\begin{eqnarray} Z &=& {\rm Tr} e^{-\beta (H_{rem}- \Gamma J_x)} \\ \nonumber
 &\approx& 
 \min\(({\rm Tr}e^{-N\beta f_{REM}},{\rm Tr} e^{-\beta Nf_{para}}\))
\end{eqnarray}
so that the equilibrium free energy is simply the minimal between the
REM and the paramagnetic one, and a first order transition appears
between the two. This allows us to immediately deduce the
thermodynamic of the problem, shown in FIG.\ref{phase}.  The first
order transition between the two phases amounts to a sudden
localization of the wave function into a sub-exponential fraction of
classical states at low $\Gamma$.

At the phase transition the two lowest levels have an avoided
crossing.  Our task is to find how close in energy they get. To
calculate the gap, we proceed as follows: Suppose that we have a value
of $\Gamma$ such that {\em for that sample} the ground state $E_o$ of
$H_{rem}$ and $-\Gamma N$ of $- \Gamma J_x$ are degenerate. The
corresponding eigenstates we denote $|SG\rangle$ and $|QP\rangle$
(spin-glass and quantum paramagnet, respectively).  To lift the
degeneracy, we diagonalize the total Hamiltonian in the corresponding
two-dimensional space and obtain
\begin{equation} H |\phi \rangle = {\Large [}\;E_o |SG\rangle\langle
  SG| - \Gamma N |QP \rangle\langle QP| \; {\Large ]}|\phi \rangle =
  \lambda |\phi \rangle \end{equation}
Multiplying this equation by $\langle SG|$ and $\langle QP|$,
respectively: \begin{eqnarray} -\lambda \langle SG|\phi \rangle &=&
  E_o \langle SG|SG \rangle \langle SG|\phi \rangle
  - \Gamma  \langle SG|X \rangle \langle SG|\phi \rangle \nonumber \\
  -\lambda \langle QP|\phi \rangle &=& E_o \langle QP|SG \rangle
  \langle SG|\phi \rangle - \Gamma \langle QP|QP \rangle \langle
  QP|\phi \rangle
\end{eqnarray} 
In order for $\lambda $ to be an eigenvalue, the determinant of this
system must vanish
\begin{equation}
  -(E_o-\lambda)(\Gamma + \lambda) + E_o \Gamma \langle SG|QP\rangle^2=0
\end{equation}
where we have used that the states are normalized. The gap is the
difference of the two solutions and reads
\begin{equation}
  gap(N,\Gamma)^2= (\Gamma-E_o)^2- 4\left[ -E_o    \Gamma + E_o \Gamma \langle SG|QP\rangle^2 \right]
\end{equation}
and at its minimum with respect to $\Gamma$, we thus get
\begin{equation}
  gap_{min}(N)=2|E_o|2^{-N/2}
\end{equation}

\begin{figure}
\includegraphics[width=7cm]{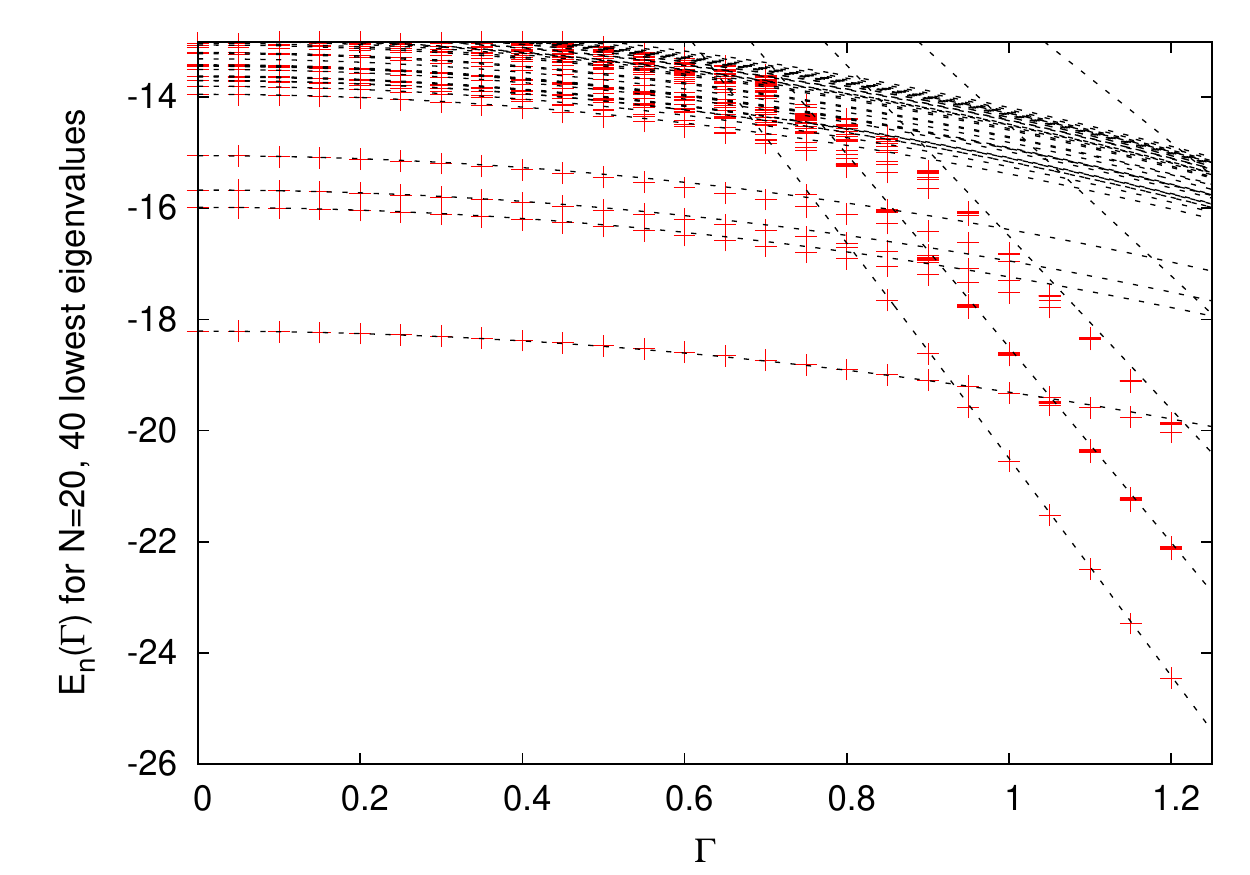}
\includegraphics[width=7cm]{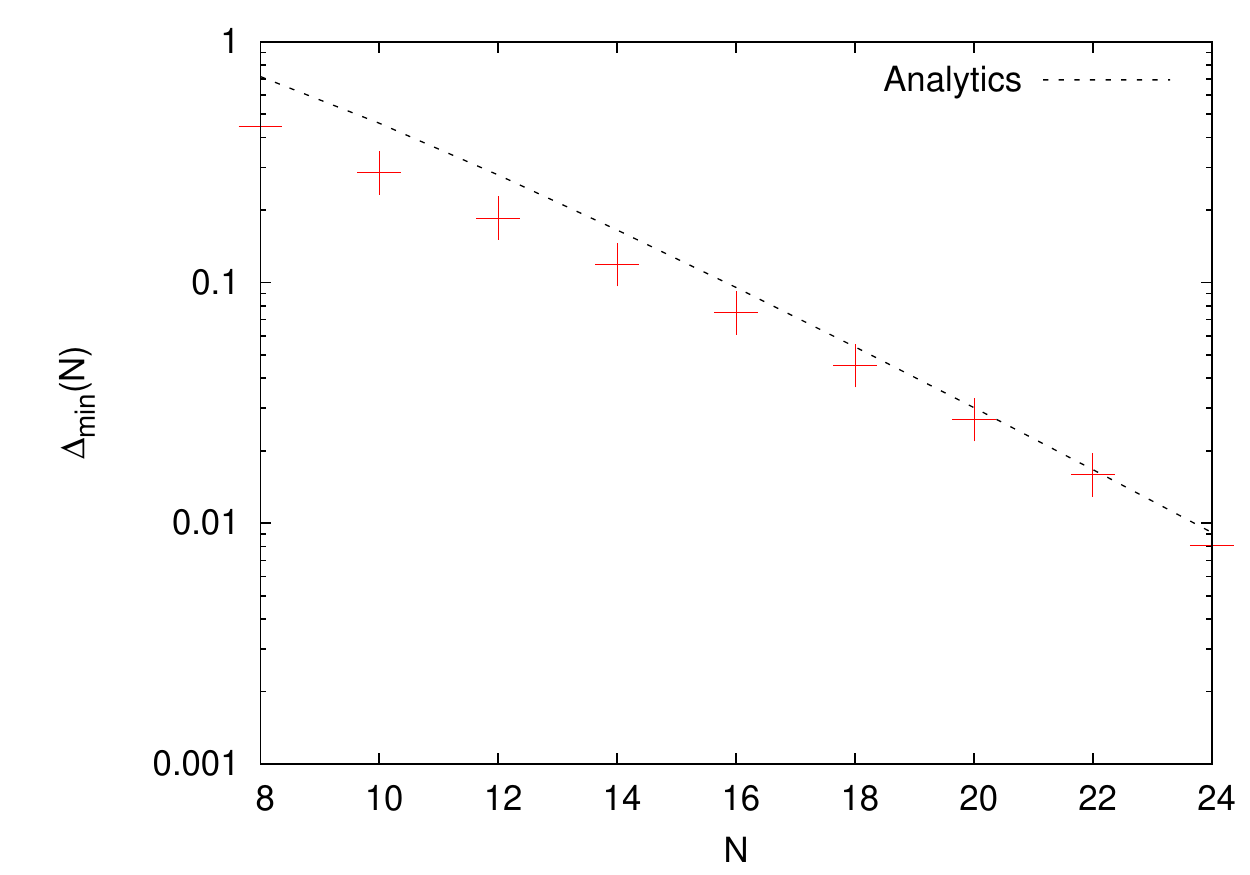}
\caption{{\bf Left:} Evolution of lowest energy levels for a single
  realization of the QREM with $N=20$ spins (dots) compared with
  analytical predictions (lines). The $40$ lower eigenvalues are
  evaluated via Lanczos iteration. {\bf Right}: Evolution of the
  ensemble averaged minimal Gap at the transition, together with the
  asymptotic analytical prediction.
  \label{levels}
}
\end{figure}

\section{Numerical simulations}
\begin{figure}
  \includegraphics[width=7cm] {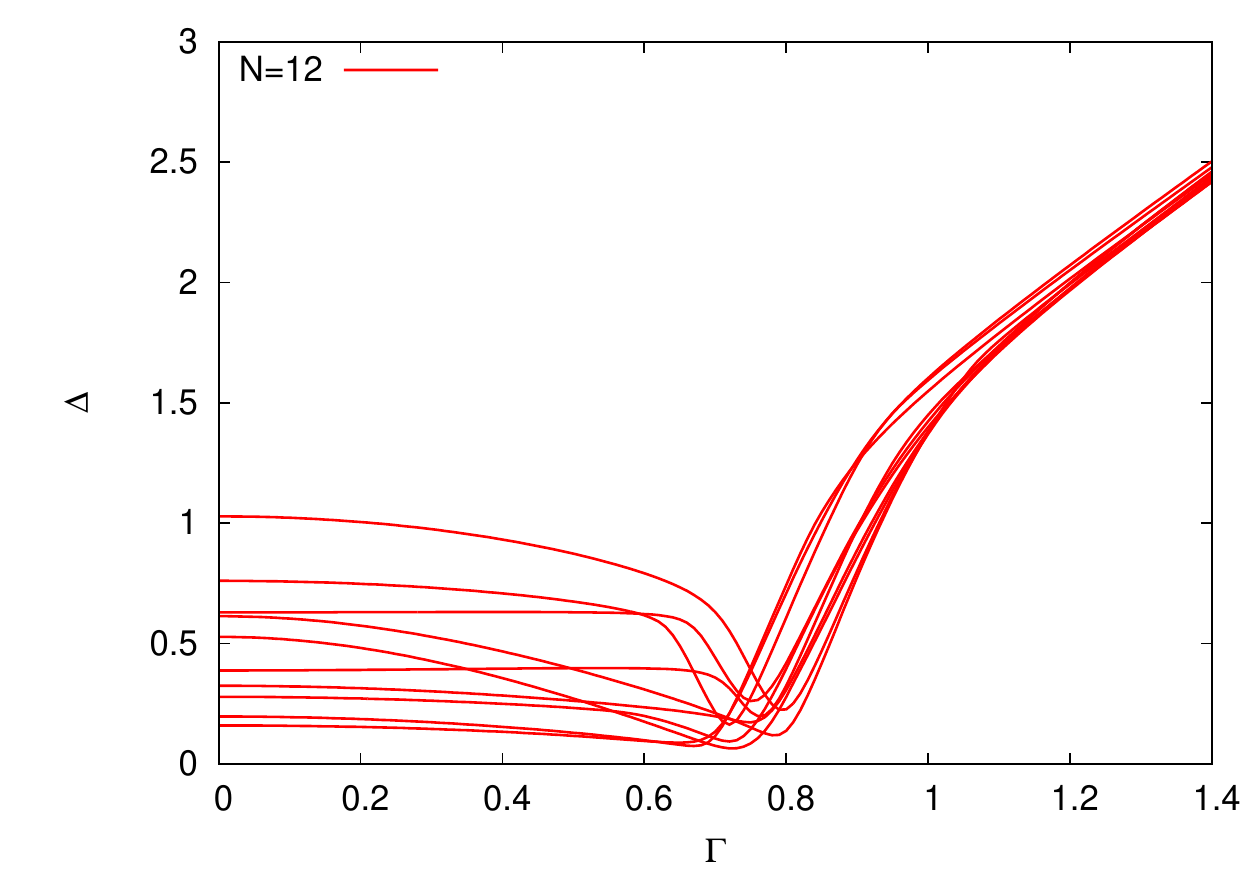}
  \includegraphics[width=7cm]{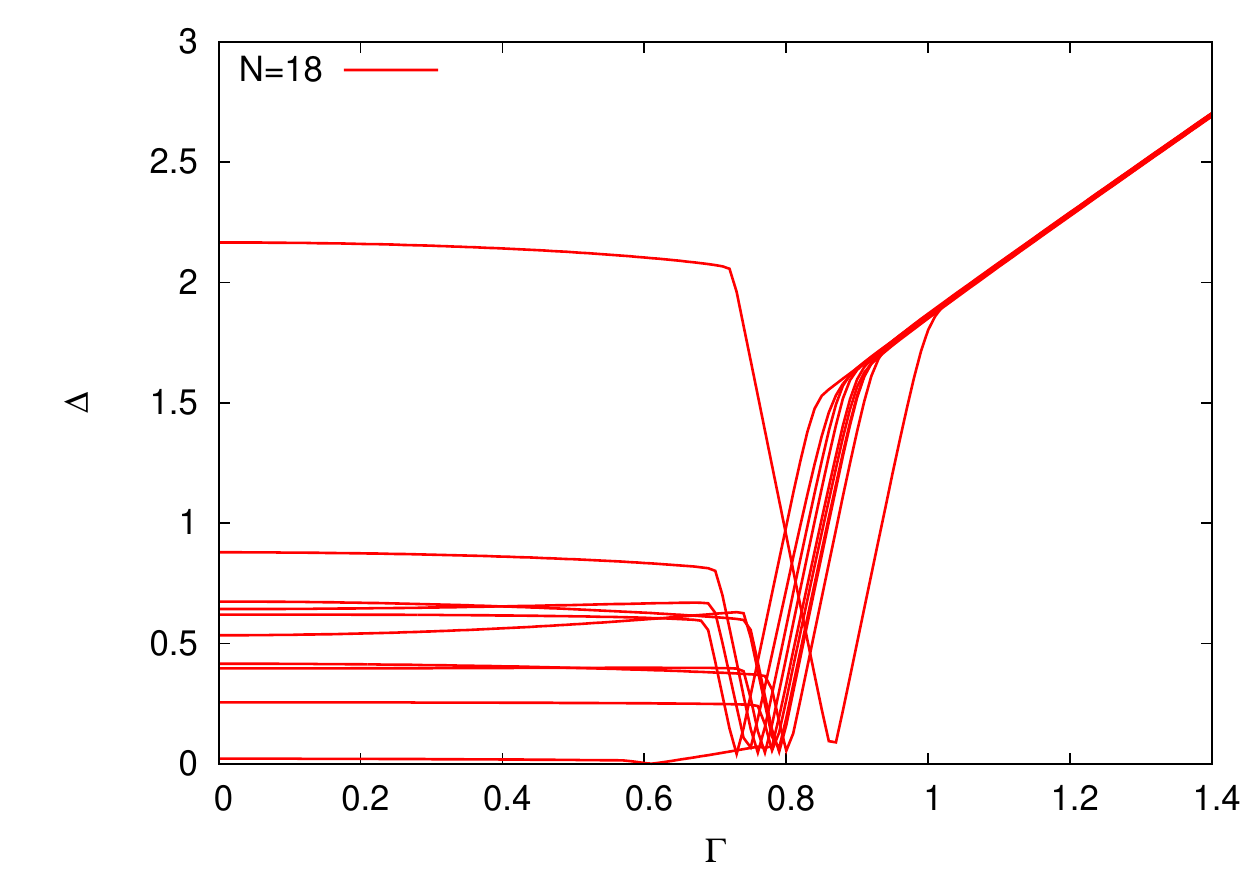}
\caption{Energy gaps between the ground state and the first excited
  state in the QREM for $10$ realizations for $N=12$ ({\bf left}) and
  $N=18$ ({\bf right}). Notice the strong fluctuations between
  samples. As $N$ grows, the region where the gap closes shrinks and
  the slope around the critical point is growing dramatically. This shows that one has to be extremely cautious when performing simulation in this region.
  \label{gaps} }
\end{figure}

The matrix elements of ${\cal{H}}_{0}^{\alpha \alpha}=E_{\alpha}$ and
${\cal{H}}_{0}^{\alpha\neq \beta}= 0 $, while $V_{\alpha \beta} = 1$
if $\alpha$ and $\beta$ are two configurations that differ by a single
spin flip and zero otherwise. ${\cal{H}}$ is sparse and can be studied
numerically rather efficiently even for large system sizes using
Arnoldi and Ritz methods \cite{Ritz}. We thus have performed a set of
numerical studies on these matrices, until $N=24$. We shows our
results in FIG.\ref{levels} and FIG.\ref{gaps}.

The prediction for the closure of the gap as
$\Delta_{min}(N)=2|E_o|2^{-N/2} $ compares very well with numerical
simulation, even for small numbers of spins (see
FIG.\ref{levels}.). Notice, however, that there are strong
fluctuations from sample to sample (see FIG.\ref{gaps}).

\section{Computing the gap in the generic case within the replica
  method}
The usual way of computing the free energy of a quantum spin system is to start with
the trace $Tr [e^{-\beta H}]$.  One works in imaginary time and uses a
a Suzuki-Trotter decomposition
\cite{Goldschmidt,Felix,Leticia,Daniel}.  In this way one obtains an
ansatz for the spin-glass phase, and a different one for the quantum
paramagnetic phase, yielding free energies $F_{SG}$ and $F_{QP}$,
respectively through $ -\beta F_{SG} \sim \ln Tr [e^{-\beta H}]$ or
$-\beta F_{QP} \sim \ln Tr [e^{-\beta H}]$, depending on the phase.

Consider now the situation with the system near the phase transition,
having two almost degenerate ground states $|SG\rangle$ and $|QP
\rangle$, $H_{SG} = \langle SG|H|SG \rangle \sim H_{QP} = \langle
QP|H|QP \rangle$ and a small interaction matrix element $H_{I} =
\langle QP|H|SG \rangle$.  To obtain the gap between levels, we have
already diagonalized this $2\times 2$ matrix exactly in the previous
section.  Alternatively, we can use the classical expansion:
\begin{equation}
  Tr [e^{-\beta H}] = \sum_r \frac{1}{r!} 
  \int dt_1 ...dt_r 
  e^{-\left[t_{tot}^{SG} H_{SG}+t_{tot}^{QP} H_{QP}\right]} 
  \; \{H_{I}\}^r
  \label{serie}
\end{equation}
where the system jumps at times $t_1,...,t_r$ between the states
$|SG\rangle$ and $|QP\rangle$, $t_{tot}^{SG}$ and $t_{tot}^{QP}=\beta
- t_{tot}^{SG}$ denote the total time spent in each state.

What equation (\ref{serie}) tells us is that if we find a solution in
imaginary time that interpolates between the quantum paramagnet and
the spin-glass solutions by jumping $r$ times, $t_1,...,t_r$, and we
find that the free energy has a value:
\begin{equation}
  \beta F_{\{r \; jumps\}}
  \sim  t_{tot}^{SG} F_{SG} +t_{tot}^{QP} 
  F_{QP}-r 
  G({\mbox{interface}})
  \label{simple}
\end{equation}
then by simple comparison $\ln H_{I} \sim G({\mbox{interface}})$
leading to ${\mbox{gap}}\sim e^G$.  An extensive value of $G$ for the
interface implies an exponentially small matrix element, and hence an
exponentially small gap. We shall now show that this is the situation
for all models of the `Random First Order' kind.  Indeed, this is just
the usual instanton construction, the difference here will be that
they take a rather unusual two-time form in problems with disorder.
 
Let us look for solutions interpolating between vacua. We consider the
p-spin model in transverse field and shall follow the presentation and
notation\footnote{With the only exception of our $q^d_{tt'}$ and
  $\tilde{q}^d_{tt'}$, which correspond to their $R_{tt'}$ and
  $\tilde{R}_{tt'}$} of Obuchi, Nishimori and Sherrington
\cite{Obuchi}.  The Hamiltonian is:
\bea {\cal H}&=&-\sum_{i_1,\ldots,i_p} J_{i_1,\ldots,i_p}
\s^z_{i_1}\ldots \s^z_{i_p} -\Gamma \sum_i \s_i^x \\
&&{\rm with}~~~ P(J_{i_1,\ldots,i_p})=\((\frac{N^{p-1}}{\pi
  p!}\))^{1/2} e^{-\frac{J_{i_1,\ldots,i_p}^2}{p!}N^{p-1}} \eea

In order to solve the system, we first apply the Trotter decomposition
in order to reduce the problem to a classical one with an additional
``time'' dimension:
\begin{equation}
  Z=\lim_{M \rightarrow \infty} {\rm Tr} \left({\rm e}^{-\beta T/M} {\rm e}^{-\beta V/M}
  \right)^M=\lim_{M \rightarrow \infty} Z_{M},
\end{equation}
where
\begin{equation}
  Z_{M}=C^{MN}{\rm Tr} \exp \left( \frac{
      \beta }{M}\sum_{t=1}^{M}\sum_{i_{1}<\ldots <i_{p}}J_{i_{1}\ldots
      i_{p}} \sigma_{i_{1},t} \ldots \sigma_{i_{p},t} + B
    \sum_{t=1}^{M}\sum_{i}\sigma_{i,t}\sigma_{i,t+1} \right),
\end{equation}
and we have introduced the constants $ B = \frac{1}{2}\ln
\coth\frac{\beta \Gamma}{M}$ and $ C=\left(
  \frac{1}{2}\sinh\frac{2\beta \Gamma}{M} \right)^{\frac{1}{2}}$.

Replicating $n$ times, and carrying over the averages over the
interactions, we get:
\begin{eqnarray}
  \left[ Z^{n}_{M} \right] \propto
  {\rm Tr} \exp
  \Biggl( \frac{\beta^2 J^2 N}{4M^2}\sum_{t,t'=1}^{M} \sum_{\mu, \nu =1}^{n}
  \left(\frac{1}{N}\sum_{i}\sigma_{i,t}^{\mu} \sigma_{i,t'}^{\nu} \right)^p
  + B\sum_{t=1}^{M} \sum_{\mu=1}^{n}\sum_{i}\sigma_{i,t}^{\mu}\sigma_{i,t+1}^{\mu}
  \Biggr), \nonumber
\end{eqnarray}
where the replica indices are denoted by $\mu$ and $\nu$.

In order to solve problem, it is then necessary to introduce a
time-dependent order parameter $q_{tt'}^{\mu\nu}$ and its conjugate
Lagrange multiplier ${\widetilde q_{tt'}^{\mu\nu} }$ for the
constraint $q_{tt'}^{\mu\nu} =\sum_{i}\sigma_{i,t}^{\mu}
\sigma_{i,t'}^{\nu}/N $. With these notations, the replicated trace
reads:
\begin{eqnarray}
  &&\left[ Z^{n}_{M} \right] = e^{  -N \beta F} = \int \prod_{\mu < \nu}\prod_{t,t'}
  dq_{tt'}^{\mu\nu} d{\widetilde q_{tt'}^{\mu\nu}}
  \prod_{\mu}\prod_{t \neq t'} dq_{tt'}^{\mu\mu} d{\widetilde
    q_{tt'}^{\mu\mu} } \nonumber \\ && \times \exp N \Biggl\{
  \sum_{t,t'} \sum_{\mu < \nu } \left( \frac{\beta^2 J^2 }{2
      M^2}\left( q_{tt'}^{\mu\nu} \right)^p - \frac{1}{M^2} {\widetilde
      q_{tt'}^{\mu\nu}}q_{tt'}^{\mu\nu} \right) \nonumber \\ &&
  + \sum_{t,t'} \sum_{\mu }
  \left( \frac{\beta^2 J^2 }{4 M^2} \left( q_{tt'}^{\mu\mu} \right)^p
    - \frac{1}{M^2} { \widetilde q_{tt'}^{\mu\mu}}q_{tt'}^{\mu\mu}
  \right) + W_o \Biggr\} 
\end{eqnarray}
where $W_o= \log {\rm Tr} \exp \left( -H_{{\rm eff}} \right) $, and:
\begin{eqnarray}
  && H_{{\rm eff}}=
  - B \sum_{t} \sum_{\mu}  \sigma_{t}^{ \mu }\sigma_{ t+1 }^{ \mu }
  - \frac{1}{M^2} \sum_{\mu < \nu}\sum_{t,t'} { \widetilde
    q^{\mu\nu}_{tt'} } \sigma_{t}^{\mu}\sigma_{t'}^{\nu} \nonumber \\
  &&     - \frac{1}{M^2} \sum_{\mu} \sum_{t \neq t'} { \widetilde
    q^{\mu\mu}_{tt'} } \sigma_{t}^{\mu}\sigma_{t'}^{\mu}.
  \label{Heff}
\end{eqnarray}
We calculate the free energy of the replicated system in the
thermodynamic limit by the saddle-point method:
\begin{eqnarray}
  q_{tt'}^{\mu \nu}=\langle \sigma_{t}^{\mu}\sigma_{t'}^{\nu} \rangle \
  &,& \ {\widetilde q_{tt'}^{\mu\nu} } = \frac{1}{2}\beta^2 J^2 p(
  q_{tt'}^{\mu \nu} )^{p-1},\label{eq:state primary cq} \nonumber \\
  q_{tt'}^{\mu \mu}=\langle \sigma_{t}^{\mu}\sigma_{t'}^{\mu} \rangle \
  &,& \ {\widetilde q_{tt'}^{\mu\mu} } = \frac{1}{4}\beta^2 J^2 p(
  q_{tt'}^{\mu \mu} )^{p-1}
  \label{equations}
\end{eqnarray}
The brackets $\langle \cdots \rangle $ denote the average with the
weight $\exp(-H_{{\rm eff}})$.

\begin{figure}[h]
  \begin{center}
    \includegraphics[width=8cm]{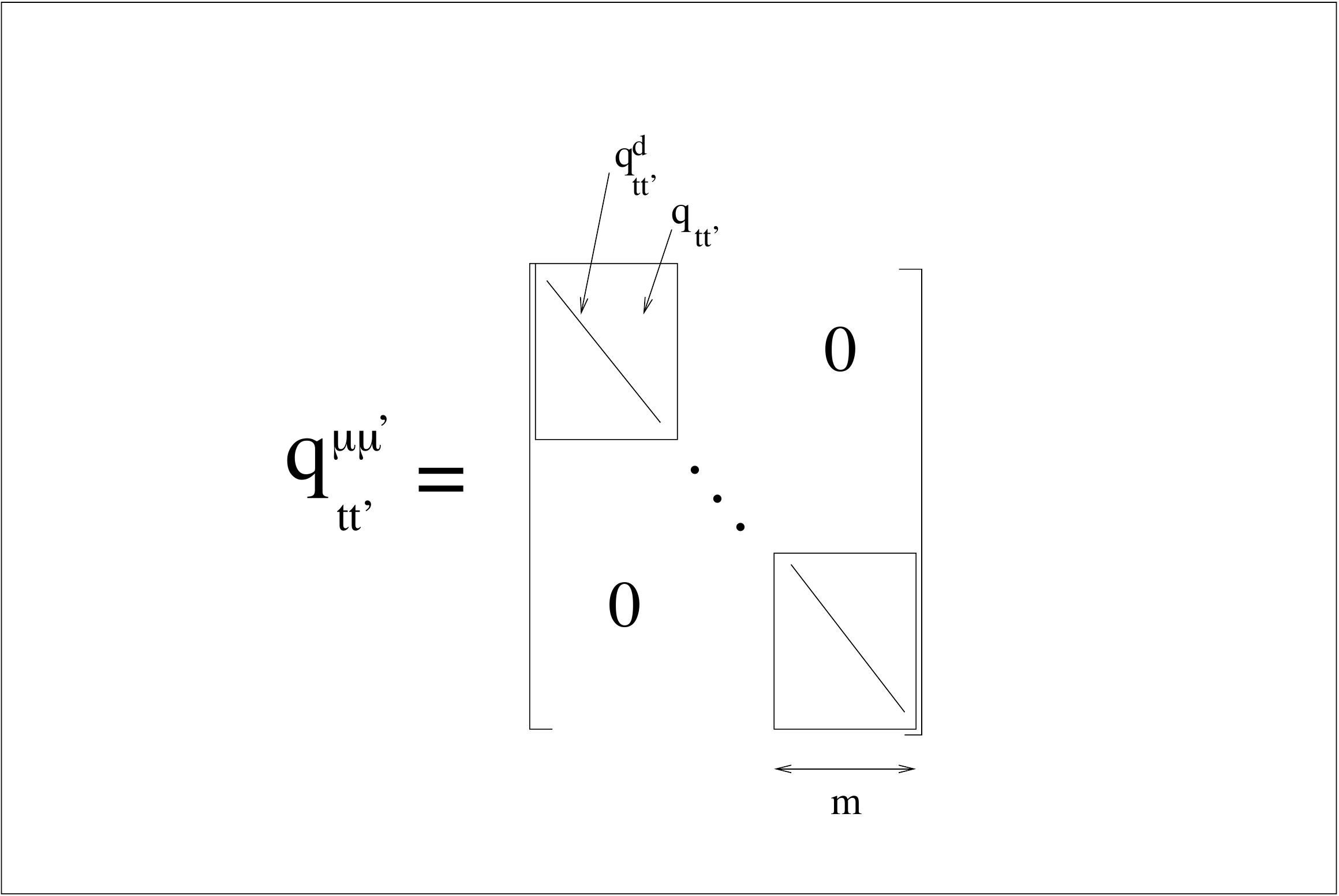}
    \caption{The one-step replica symmetry breaking ansatz. The
      two-time two-replica overlap $q_{tt'}^{\mu\nu}$ has the
      following structure. We divide replica in m groups and the
      overlap between replica in different group is zero while the
      overlap between replicas of the same group is $q_{tt'}$ if the
      replica are different ($/mu \neq \nu$), are $q^d_{tt'}$ is the
      replica are the same ($\mu=\nu$ on the diagonal). }
    \label{rsbr}
  \end{center}
\end{figure}

We now make the one-step replica symmetry broken ansatz: for every
$(t,t')$ the parameter $q_{tt'}^{\mu\nu} $ is as in Fig \ref{rsbr}.
Within the 1RSB ansatz, the replicated free energy reads:
\begin{eqnarray}
  -\beta f &=&\int dt \; dt' \;\left\{
    -\frac{\beta^2 J^2}{4} (1-m)  q^p_{tt'} \right.
  \nonumber \\
  & &+ \left. \frac{1}{2} (1-m)
    \tilde q_{tt'}   q_{tt'} +\frac{\beta^2 J^2}{4}  [q^d_{tt'}]^p
    -   \tilde q^d_{tt'}   q^d_{tt'} \right\}- W_o[q^d_{t't'},q_{tt'}]
  \label{ooo}
\end{eqnarray}

Up to here we have followed the standard
steps~\cite{Goldschmidt,Felix,Leticia,Daniel,Obuchi} Solutions within
this ansatz corresponding to the spin glass and the quantum paramagnet
have been studied by several authors
\cite{Goldschmidt,Felix,Leticia,Daniel,Obuchi}. It turns out that the
phenomenology for finite the $p$ is very close to that of the QREM.

\begin{figure}[ht]
  \begin{center}
    \includegraphics[width=6cm]{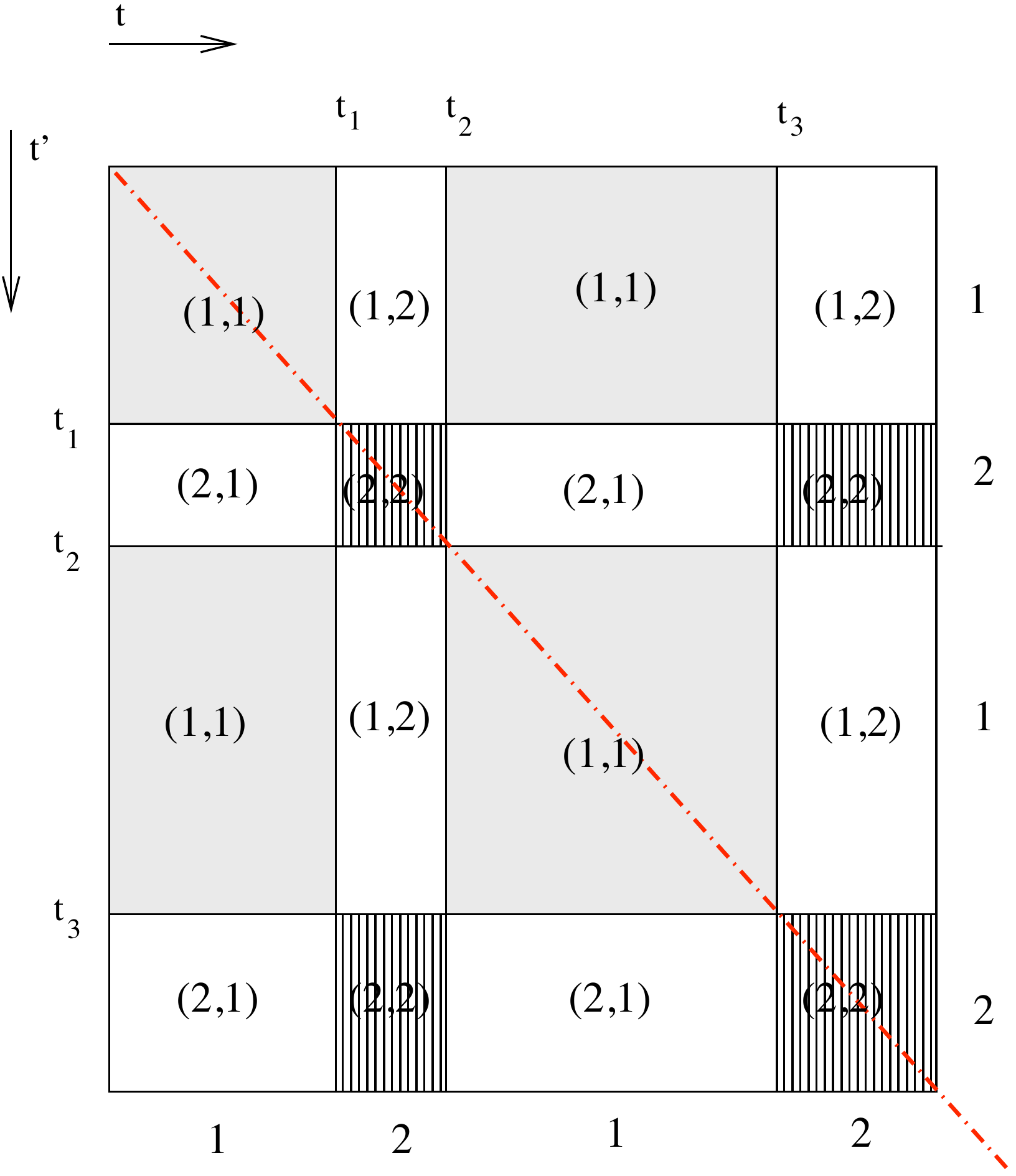}
  \end{center}
  \caption{A multi-instanton configuration for $q^d_{t,t'}$,
    $q_{t,t'}$, ${\tilde q}^d_{t,t'}$ and $\tilde{q}_{t,t'}$.  Three
    different shadings correspond situations in which the system is in
    the glass phase at both times, the paramagnetic phase at both
    times and in a different phase at each time.  In general, the
    functions $q_{t,t'}$, $\tilde{q}_{t,t'}$ are approximately
    constant well within each region, with smooth, finite width
    interfaces. The diagonal elements $q^d_{t,t'}$ and ${\tilde
      q}^d_{t,t'}$ have an additional crest along the diagonal
    $t=t'$. In the large $p$ limit, we may neglect this crest, in the
    so-called `static approximation'.  In addition ${\tilde
      q}^d_{t,t'}$ and $\tilde{q}_{t,t'}$ become either infinity or
    zero, with sharp interfaces: this allows for a complete solution
    at large $p$.  }
  \label{fig2}
\end{figure}

We now part company with the previous literature, and consider a
solution corresponding to the high-$\Gamma$ phase in the interval
$(0,t_1)$, $(t_2,t_3)$,..., that jumps to the low-$\Gamma$ phase where
it stays in the intervals $(t_1,t_2)$, $(t_4,t_5)$,....  The
time-dependence of $q_{t,t'}$ and $q^d_{t,t'}$ are is as in Fig.
(\ref{fig2}).

\subsection{Recovering the REM result}

Let us now specialize to the large-$p$ situation, and we shall later
indicate how to proceed in general.  Following Obuchi {\it et al.}, we
note the saddle point equations imply that for large $p$ either
$(q_{tt'},q^d_{tt'}, \tilde q_{tt'}, \tilde
q^d_{tt'})=(1,1,\infty,\infty)$ or $(q_{tt'},q^d_{tt'}, \tilde
q_{tt'}, \tilde q^d_{tt'})=(<1,<1,0,0))$.  This implies that the form
of the instanton configuration of $\tilde q^d_{tt'}$ and $\tilde
q_{tt'}$ is the same as the one of $(q_{tt'},q^d_{tt'}$ but with the
values jumping from zero to infinity.  If in addition we make the
`Static approximation' that as applied to this case consists of
assuming that inside each block the order parameters $q^d$ and $\tilde
q^d$ are constant, we conclude that for all times we can write the
`tilde' variables as:
\begin{equation}
  2\tilde q^d_{t't'} -\tilde q_{t't'} = r^d_t r^d_{t'} \qquad ; 
  \qquad \tilde q_{t't'} = r_t r_{t'}
\end{equation}
where $r_t$ and $r^d_{t}$ are large in the time intervals when the
system is in the glass state, and drop to zero when it is not. (The
solutions in the literature correspond to time-independent $r$, taking
a value corresponding to either the glass or the paramagnetic phase)

Because $ q^d_{t't'}, q_{t't'}$ are either zero or one, we can write:
\begin{eqnarray}
  \int dt \; dt' \; q_{tt'} \tilde q_{tt'} &\sim& \left[\int dt \; r(t)\right]^2
  = I^2 \nonumber \\ 2 \int dt \; dt' \;\tilde q_{tt'} \tilde q^d_{tt'} &=& \int dt \;
  dt' \; [r^d(t)r^d(t') + r(t)r(t')]= I_d^2+ I^2
\end{eqnarray}
with the definitions $I \equiv \int dt \; r(t)$ and $ I_d \equiv \int
dt \; r^d(t)$.

We further decouple the replicas in the single-spin term in the usual
way:
\begin{eqnarray}
  W_o&=& \log {\rm Tr} \exp \left( -H_{{\rm eff}} \right) =
  -\frac{1}{m} \log \left\{ \int Dz_2 \left[ \int Dz_3 \;  \right.\right.
  \nonumber \\ & & \left. \left. 
      {\rm Tr} \left(e^{ -B\sum_t \sigma_t \sigma_{t+1}-
          \frac{1}{M} \sum_t (z_2 r_t+ z_3 r^d_t) \sigma_t
        }\right) \right]^m \right\}
  \label{pp}
\end{eqnarray}
We can go back to the quantum representation, to write the trace in
(\ref{pp}) as a single quantum-spin in a time-dependent field:
\begin{equation}
  {\rm Tr} \left( {\cal{T}} 
    e^{\int dt' (A(t') \sigma^z + \beta \Gamma \sigma^x)} \right)
\end{equation}
where $ {\cal{T}}$ denotes time-order (a necessity here because of the
time-dependence in the exponent) , and $ A(t) \equiv ( z_3 r^d_t+z_2
r_t)$.  In the periods where $A$ is non zero, it has a large real part
and the spins are polarized along the $z$ direction. During the
periods when $r_t=r^d_t=A=0$ only the transverse term proportional to
$\Gamma $ plays a role. This corresponds to the switching between
spin-glass and quantum paramagnetic phases, respectively. The trace
can be computed easily by changing bases accordingly.  At low
temperatures, the `field' in the $x$ direction $\beta \Gamma$ is
strong, while the field in the $z$ direction $|A(t)|$ is either zero
or $|A(t)|>>\beta \Gamma$ (for large $p$, cf. Eq.  (\ref{equations})).
We conclude then that only one polarization of the spin contributes in
each time interval.  Denoting
\begin{eqnarray}
  && {\rm Tr} \left( e^{\int dt' (A(t') \sigma^z + \beta \Gamma
      \sigma^x)} \right) \sim \langle z_1|
  e^{\int_{t_n}^{t_f} dt' A(t') \sigma^z } |z_1\rangle \langle
  z_1|x\rangle  \nonumber \\
  &&
  \langle x| e^{\beta (t_n-t_{n-1}) \Gamma \sigma^x}
  |x\rangle \; \langle x|z_2\rangle \; \dots \; \langle z_n|
  e^{\int_{t_1}^{0} dt' A(t') \sigma^z} |z_n\rangle
\end{eqnarray}
Where we have defined $|x\rangle$ the lowest eigenvalue of $\sigma^x$,
and the $|z_i\rangle$ are either the lowest or the highest eigenvalue
of $\sigma^z$, depending on the sign of $A(t)$ (which in turn depends
on $z_2$,$z_3$) during that interval. Because, as mentioned above, at
each time only one field dominates, and the way we chose the
$|z_i\rangle$, we have:
\begin{equation}
  \langle z_i| e^{\int_{t_n}^{t_f} dt' A(t') \sigma^z } |z_i\rangle
  \sim e^{\int_{t_n}^{t_f} dt' |A(t')|}
\end{equation}
and
\begin{equation}
  \langle x| e^{ \beta \Gamma \sigma^x } |x\rangle \sim e^{\beta \Gamma
  }
\end{equation}
Hence, if we denote $t^{SG}=\Theta \beta$ the total time in which
$r_t$, $r^d_t$ are not zero, and $q_t=q^d_t=1$, and $t^{QP}=(1-\Theta)
\beta$ the rest, the action becomes:
\begin{eqnarray}
  -\beta f &=&\Theta^2 \;\left\{ -\frac{\beta^2 J^2}{4} (1-m)  
    +\frac{\beta^2 J^2}{4}   \right\} - \frac{1}{2} I_d^2 -
  \frac{m}{2} I^2  W_z \nonumber \\ &+& (1-\Theta) \beta \Gamma +
  {\mbox{(numb. of jumps)}} \times \ln |\langle x|z \rangle|
  \label{ppp}
\end{eqnarray}
where
\begin{eqnarray}
  W_z&=& -\frac{1}{m} \log \left\{ \int Dz_2 \left[ \int Dz_3 \;
      e^{\int dt|z_2  r_t +  z_3 r^d_t| }
    \right]^m \right\} \nonumber \\&=&  
  -\frac{1}{m} \log \left\{ \int Dz_2 \left[ \int Dz_3 \;
      e^{|z_2  I +  z_3 I_d | } \right]^m \right\}\nonumber
\end{eqnarray}
This can be evaluated by saddle point \cite{Goldschmidt,Obuchi},
putting $z_3=I_d y_3$ and $z_2=I y_2$ and recognizing that $I,I_d$ are
large. There are two saddle points: $(y_2=1,y_3=1)$ and
$(y_2=-1,y_3=-1)$ with the same contribution. A short calculation
yields:
\begin{equation}
  W_z \sim  \frac{1}{2} I_d^2 + \frac{m}{2} I^2 + \ln(2) 
\end{equation}
Equation (\ref{ppp}) becomes
\begin{equation}
  -\beta f = 
  m \Theta^2 \frac{\beta^2 J^2}{4}   
  - \frac{\ln (2)}{m} +  (1-\Theta) \beta \Gamma +
  {\mbox{(numb. of jumps)}} \times \ln |\langle x|z \rangle|
  \label{ooo1}
\end{equation}
Taking saddle point with respect to $m$ gives
$m=\frac{2\sqrt{2}}{\Theta \beta J}$.  We finally obtain
\begin{equation}
  -\beta f = 
  \Theta \sqrt{\ln (2)} \frac{  \beta J}{2}   
  +  (1-\Theta) \beta \Gamma +
  {\mbox{(numb. of jumps)}} \times \ln |\langle x|z \rangle|
  \label{ooo2}
\end{equation}
This formula is of the form (\ref{simple}).  It gives the contribution
to $Tr[e^{-\beta H}]$ of the process with a number of jumps spending a
fraction $\Theta $ in the glass state and $(1-\Theta)$ in the
paramagnetic state. {\em We have hence showed that the logarithm of
  matrix element is indeed the single-spin element $\sim N \ln
  |\langle x|z \rangle|= -N \ln(2)$ from which we recover the gap}
$\sim 2^{-N}$ obtained previously.

\subsection{The general case}

In the general case, we need to determine the functions $q^d_{tt'}$
and $q_{tt'}$ for the interval $0<t,t'<\beta$. It suffices to compute
a single jump solution, and to calculate the free energy cost of the
"wall" in the two times. This may be done numerically, discretizing
the times (i.e. working in a two-time grid) and {\em minimizing} with
respect to $q_{tt'}$ while {\em maximizing} with respect to the
diagonal replica parameters $q^{d}_{tt'}$, as is usual in the replica
trick.

The solution must be such that the order parameters $q^d_{tt'}$ and
$q_{tt'}$ are for small $t,t'$ close to those computed for the quantum
paramagnet, for $t,t'$ near $\beta$ close to the solution for the
glass phase, and for mixed times (one large and one small) a constant
corresponding to the phase-space overlap between both phases ---
typically zero.
 
At precisely the values of parameters such that the free energies of
both pure solutions coincide, we have an extra free energy density
cost is due exclusively the (smooth) walls separating the four
quadrants in the two-time plane.  Their precise form, and their free
energy cost, has to be determined numerically.  If their cost in free
energy density is larger than zero, this is the value of the
exponential dependence in the gap.
 
An interesting situation arises in models where the one-step replica
symmetry breaking ansatz for the pure quantum phases is not exact,
such as the quantum Sherrington-Kirkpatrick model.  In that case, we
should generalize the ansatz to allow a full replica symmetry
breaking. This will not be the end of the story in such models, as the
cost of the wall in the two-time plane may vanish to leading order in
$N$, reflecting the fact that barriers are subextensive in height and
equilibrium states are at all possible distances.  In such cases, one
should compute the free energy of an instanton solution scaling slower
than the system size, i.e. the subextensive corrections.  This is
beyond our present technical capabilities.

 \section{Discussion }
 We have described an analytic technique to compute the minimal gap
 between the two lower quantum levels of a class of mean-field glass
 problems. The nature of the solution suggests that all models having
 the "Random First Order" phenomenology will have an exponentially
 small gap, implying that quantum annealing cannot find the ground
 state in subexponential time.  The same two-time instanton technique
 may be generalized to "dilute" systems, where the connectivity is
 finite but locally tree-like and it would thus be interesting to
 adapt it to the cavity setting developed in \cite{Cavity}.  The
 reader interested in seeing how the computation works in a simpler
 setting, in absence of disorder, is referred to \cite{ferro}.

 Notice also that over the last few months, the first order transition
 scenario has been confirmed by numerical simulations in
 \cite{youngnew} as well as with analytical results in diluted systems
 \cite{preparation}.

%\appendix
%\section{First Appendix} %Empty argument \section{} yields `Appendix'. 
%
%\section{Second Appendix}


\begin{thebibliography}{99}
\bibitem{foot2} Within each {degenerate} subspace the perturbation
  ${\cal{H}}_0$ acts as a random orthogonal matrix of variance
  $N2^{-N}$. This is tiny and weakly lifts the degeneracy.
\end{thebibliography}
\end{document}